\documentclass{nature}
\usepackage{multicol}
\usepackage{graphicx}	
\usepackage{amsmath}
\usepackage{amssymb}
\usepackage{lineno}

\title{A dusty star-forming galaxy at {\textbf{\textit z}}=6 revealed by strong gravitational lensing}

\author{Jorge A. Zavala$^{1,2*}$, Alfredo Monta\~na$^{3}$, David H. Hughes$^1$, Min S. Yun$^4$,  R.\,J.~Ivison$^{5,6}$, Elisabetta Valiante$^7$,
David Wilner$^8$, Justin Spilker$^9$, Itziar Aretxaga$^1$, Stephen Eales$^7$, Vladimir Avila-Reese$^{10}$, Miguel Ch\'avez$^1$, Asantha Cooray$^{11}$, 
Helmut Dannerbauer$^{12,13}$, James S. Dunlop$^6$, Loretta Dunne$^{6,7}$, Arturo I. G\'omez-Ruiz$^{3}$, Micha{\l} J. Micha{\l}owski$^{14}$, Gopal Narayanan$^4$, 
Hooshang Nayyeri$^{11}$, Ivan Oteo$^{6,5}$, Daniel Rosa Gonz\'alez$^1$, David S\'anchez-Arg\"uelles$^1$, F. Peter Schloerb$^4$, Stephen Serjeant$^{15}$, Matthew W. L. Smith$^7$, 
Elena Terlevich$^1$, Olga Vega$^1$, Alan Villalba$^1$, Paul van der Werf$^{16}$, Grant W. Wilson$^4$,  Milagros Zeballos$^1$}
\begin{document}
\maketitle
\begin{affiliations}
{\footnotesize
 \item Instituto Nacional de Astrof\'{i}sica, \'{O}ptica y Electr\'{o}nica (INAOE), Luis Enrique Erro 1, 72840, Puebla, Mexico
 \item Department of Astronomy, The University of Texas at Austin, 2515 Speedway Boulevard, Austin, TX 78712, USA 
 \item CONACyT-Instituto Nacional de Astrof\'isica, \'Optica y Electr\'onica, Luis Enrique Erro 1, 72840, Puebla, Mexico
 \item Department of Astronomy, University of Massachusetts, MA 01003, USA
 \item European Southern Observatory, Karl Schwarzschild Strasse 2, Garching, Germany
 \item Institute for Astronomy, University of Edinburgh, Royal Observatory, Blackford Hill, Edinburgh EH9 3HJ, UK
 \item School of Physics and Astronomy, Cardiff University, The Parade, Cardiff CF24 3AA, UK
 \item Harvard-Smithsonian Center for Astrophysics, 60 Garden Street, Cambridge, MA 02138, USA
 \item Steward Observatory, University of Arizona, 933 North Cherry Avenue, Tucson, AZ 85721, USA
 \item Instituto de Astronom\'ia, Universidad Nacional Aut\'onoma de M\'exico, A.P. 70-264, 04510, CDMX, Mexico
 \item Dept. of Physics \& Astronomy, University of California, Irvine, CA 92697, USA
 \item Instituto de Astrof\'isica de Canarias (IAC), E-38205 La Laguna, Tenerife, Spain  
 \item Universidad de La Laguna, Dpto. Astrof\'isica, E-38206 La Laguna, Tenerife, Spain 
 \item Astronomical Observatory Institute, Faculty of Physics, Adam Mickiewicz University, ul.~S{\l}oneczna 36, 60-286 Pozna{\'n}, Poland
 \item Department of Physical Sciences, The Open University, Milton Keynes, MK7 6AA, UK
 \item Leiden Observatory, Leiden University, P.O. Box 9513, NL-2300 RA Leiden, The Netherlands 
 }
 \end{affiliations}

 \begin{multicols}{2}
\begin{abstract}
Since their discovery, submillimeter-selected galaxies\cite{ref53,ref2} have revolutionized the 
field of galaxy formation and evolution. 
From the hundreds of square degrees mapped at
submillimeter wavelengths\cite{ref3,ref4,ref22},  
only a handful of sources have been confirmed to lie 
at $z>5$ (ref.\cite{ref7,ref8,ref9,ref10,riechers17}) and only two at $z\ge6$ (ref.\cite{ref11,ref57}).
All of these SMGs are rare examples of extreme starburst galaxies with star formation rates of
$\gtrsim 1000$\ M$_\odot$\ yr$^{-1}$ and therefore are not representative of the general population
of dusty star-forming galaxies. Consequently, our understanding of the nature of these sources, 
at the earliest epochs, is still incomplete. Here we report the spectroscopic identification of a
gravitationally amplified ($\mu=9.3\pm1.0$) dusty star-forming galaxy at $z=6.027$.
After correcting for gravitational lensing we derive an intrinsic less extreme star formation rate 
of $380\pm50$\ M$_\odot$\ yr$^{-1}$ for this source, and find that its gas and dust properties are similar
to those measured for local Ultra Luminous Infrared Galaxies, extending the local trends to a 
poorly explored territory in the early Universe. The star-formation efficiency of this galaxy is
similar to those measured in its local analogues\cite{ref17}, despite a $\sim12$ Gyr difference in cosmic time.
\end{abstract}

\noindent HATLAS\ J090045.4+004125 ($\alpha=09^\text{h}00^\text{m}45.8$, $\delta=+00^\circ41'23''$; hereafter G09 83808, since it was detected in the GAMA 09hrs field) 
is part of a sub-sample of the {\it Herschel} ATLAS `500\ $\mu$m-riser' galaxies\cite{ref12} with  ultra-red
far-infrared (FIR) colours  of  $S_{500_{\mu\rm m}}/S_{250_{\mu\rm m}} > 2$ and $S_{500_{\mu\rm m}}/S_{350_{\mu\rm m}} > 1$, with a flux density threshold 
of $S_{500_{\mu\rm m}}<80\ \rm mJy$. 
The FIR colours of this source are consistent with thermal dust emission redshifted to $z>4$ and  represent a relatively simple selection criterion to find 
high-redshift galaxies. A similar selection  allowed  the identification of HFLS3\cite{ref11}, an extreme starburst galaxy (even after corrected for 
gravitational amplification\cite{cooray14}) at $z = 6.3$, in the HerMES blank field survey\cite{ref3}.

\noindent G09 83808 was observed, among other ultrared-{\it Herschel} dusty star-forming galaxies, as part of a follow-up program with the 
Large Millimeter Telescope {\it Alfonso Serrano} (LMT)
using the AzTEC camera, in order to obtain 
higher angular resolution ($\sim8.5$\ arcsec) continuum observations at 1.1\ mm. A sub-sample of those galaxies detected as a single source in the AzTEC images 
(i.e. with no evidence of multiple components)
and with photometric redshifts of $z>4$, was selected for spectroscopic observations in the 3\ mm band using the Redshift Search Receiver (RSR) on the LMT.
\noindent In the LMT/RSR spectrum of G09 83808 we identify three emission lines corresponding to $^{12}$CO($6-5$), $^{12}$CO($5-4$), and H$_2$O($2_{11}-2_{02}$) (see Fig. 1).
Based on these lines we unambiguously determine the galaxy redshift to be $z=6.0269\pm0.0006$ (i.e. when the Universe was just 900 million years old).
Follow-up observations with The Submillimeter Array (SMA) telescope confirm this solution through the  detection of the redshifted [CII] ionized carbon line at 270.35\ GHz (see Fig. 1).

\vspace{-0.2cm}
\noindent High-angular resolution observations ($0.24\rm\ arcsec \times0.13\ arcsec$, corresponding to a physical scale of $\sim1$\ kpc at this redshift)
taken with the Atacama Large Millimeter/submillimeter Array (ALMA; see Methods section)
at $\sim890\ \mu\rm m$ reveal a double arc structure (in a partial Einstein ring configuration of radius $\sim 1.4$\ arcsec)
around a foreground galaxy at $z=0.776$ (see Fig. 2), implying 
strong gravitational amplification of the high-redshift background galaxy. 
Using these ALMA continuum observations to constrain the effects of gravitational lensing, modelling directly the visibilities in the {\it uv} plane
(see Methods section for additional details), we derive a gravitational amplification factor of $\mu=9.3\pm1.0$. 
This amplification factor is used to derive the intrinsic physical properties of G09 83808.

\vspace{-0.2cm}
\noindent Using the {\it Herschel} 250, 350, and 500~$\mu\rm m$ photometry\cite{ref12}, combined with the SCUBA-2 850\ $\mu\rm m$\cite{ref12} imaging and our AzTEC 1.1\ mm
observations (see Table 1), we model the 
continuum spectral energy distribution (SED; see Figure 3). We estimate an infrared (IR, $8-1000\ \mu\rm m$) luminosity, $L_{\rm IR}$, of $3.8\pm0.5\times10^{12}\rm\ L_\odot$ 
(corrected for gravitational magnification) which implies a dust-obscured star formation rate (SFR) of $380\pm50\rm\ M_\odot yr^{-1}$ 
(see Methods section for more information). This means that G09 83808 is a member of the Ultra Luminous Infrared Galaxy (ULIRGs\cite{ref17}) population.  
This is one of the first SMG with an unambiguous spectroscopic redshift in this luminosity range at $z\gtrsim5$, 
lying between the extreme obscured starbursts\cite{ref7,ref8,ref9,ref11,ref57} 
($\gtrsim 1000$\ M$_\odot$\ yr$^{-1}$) discovered at submm wavelengths and the UV/optical selected star-forming galaxies with follow-up detections at submm
wavelengths\cite{capak2015,ref18,ref19} ($\lesssim 100$\ M$_\odot$\ yr$^{-1}$).  Within this luminosity range only a handful of galaxies at $z>6$ are known, which 
were recently discovered around quasars\cite{ref58} thanks to the serendepitously detection of a single emission line associated to [CII].

\vspace{-0.2cm}
\noindent Although these galaxies are unreachable with the current generation of submm wide-area surveys\cite{ref3,ref4} without the 
benefit of gravitational amplification, they can be found in the deepest
surveys recently achieved  with ground-based telescopes, such as the James Clerk Maxwell Telescope (JCMT) SCUBA-2 Cosmology Legacy Survey (S2CLS). However,
none of them has yet been spectroscopically confirmed. With the caveat of using the position of the dust SED peak as an estimation of redshift,
a study based on S2CLS observations\cite{ref22} has derived a comoving space density of $3.2\times10^{-6}\ \rm Mpc^{-3}$ for sources with $300<\rm SFR < 1000\ M_\odot\ yr^{-1}$
at $5<z\lesssim6$ (i.e. in the range probed by our galaxy). With a duty-cycle correction of $\approx 40\ \rm Myr$, 
as the gas depletion time scale measured for G09 83808 (see below) and other galaxies\cite{ref11, ref56}, we estimate the corrected comoving space density 
of this population of galaxies to be  $\approx2\times10^{-5}\ \rm Mpc^{-3}$, which perfectly matches that of massive quiescent galaxies at $z\approx3-4$ (refs.\cite{nayyeri,ref24}).
This suggests, that these ULIRG-type galaxies at 
$5\lesssim z\lesssim6$ are the progenitors of these quiescent galaxies, which cannot be explained only by the rare extreme starburst galaxies (like HFLS3), 
since they are an order of magnitude less abundant\cite{ref12}.

\vspace{-0.2cm}
\noindent Based on the CO lines detected in the LMT/RSR spectrum we derive a molecular gas mass of $\rm M(H_2)=1.6\pm0.6\times10^{10}\rm\ M_\odot$ (see Methods section for 
details). This implies a gas depletion 
timescale of $\rm M(H_2)/SFR\approx40$\ Myr, consistent with the value found for other SMGs at lower redshifts with ULIRG-luminosity\cite{ref28}. 
G09 83808 shows a remarkable large gas mass fraction of $f_{\rm gas}=M_{\rm H_2}/M_{\rm dyn}\sim 60$\% (see Methods secction), among the largest measured for
star forming galaxies at $z\approx2-3$ (ref.\cite{ref59}). The CO(6-5)/CO(5-4) line luminosity ratio of $0.4\pm0.1$ 
is in agreement with local ULIRGs (although lower than the average\cite{greve14}), and implies  a CO ladder peaking at 
J$\le5$ (i.e. less excited than AGN-dominated galaxies\cite{ref51}). These two CO transitions, as well as the H$_2$O line, 
lie (within the error bars) on their respective FIR/IR-line luminosity relations ($L_{\rm FIR}\propto L_{\rm CO(6-5)}^{0.93}$, $L_{\rm FIR}\propto L_{\rm CO(5-4)}^{0.97}$, and 
$L_{\rm H_2O}\propto L_{\rm IR}^{1.16}$) found for local ULIRGs and lower redshifts SMGs\cite{greve14,yang2016}. 
The star-formation efficiency (SFE) of our galaxy, estimated through the $L'_{CO}-L_{\rm IR}$ relation (which describes the relationship
between the luminosity due to star formation and the gas content), is similar to local (U)LIRGs (see Fig. 4). Then, the same SFE
can be found across several decades of molecular gas masses from $z=6$ to $z\sim0$ (i.e., during the last 12.8 Gyr of the Universe).
In addition, the estimated dust mass of $M_d=1.9\pm0.4\times10^8\rm\ M_\odot$ results in a gas-to-dust ratio, $\delta_{\rm GDR}$, of $80\pm30$.
This is in agreement with the value estimated for HFLS3\cite{ref11} and also with local (U)LIRGs\cite{ref29} ($\delta_{\rm GDR}=120\pm28$). 

\vspace{-0.2cm}
\noindent The luminosity of the [CII] ionized carbon line detected with the SMA is $1.3\pm0.4\times10^{9}\rm\ L_\odot$ which corresponds to a [CII]/FIR ratio of 
$3.4\pm1.1\times10^{-4}$, a value that is among the lowest measured for local (U)LIRGs and SMGs. As shown in Figure 4, our source follows the same [CII]
deficiency trend measured for local LIRGS\cite{ref30} extending it to $\rm L_{FIR}\gtrsim 10^{12}\ L_\odot$ and up to $z=6$. 
The [CII]/FIR ratio of G09 83808 is also consistent with the lowest values measured for lower-redshift SMGs and lies on a region where SMGs and AGN-host galaxies converge (Fig. 4).
It may be the case that other SMGs suffer from gravitiational amplification, which could help to reduce the large scatter 
since many of these galaxies should fall along the LIRG relation when corrected for magnification. However, the intrinsic scatter in the relation is high\cite{ref30}, even for the 
local sample, and therefore, larger samples of SMGs are required to derive conclusions about the origin of the [CII] deficiency.

\vspace{-0.2cm}
\noindent We confirm the existence of ULIRG-like galaxies within the first billion years of Universe's history. These sources may be more
representative of the dusty star-forming galaxy population at these epochs than the extreme starbursts previously discovered. 
Four emission-line-selected galaxies with similar luminosities and redshifts have been recently found around quasars\cite{ref58}
(with the caveat of using just one line for redshift determination), however, the properties of these sources  may be affected by 
the companion quasar and therefore not representative of the whole population. Although G09 83808 shows similar properties to those 
measured in lower-redshift SMGs, its higher dust temperature ($T_d=49\pm3$\ K) and compact morphology ($R_{1/2}=0.6\pm0.1$\ kpc) resemble that of local ULIRGs.
For comparison, typical UV/optically-selected star-forming galaxies at $z\sim6$ have SFRs $\sim10$ times lower and radii $\sim1.7$ times larger than G09 83808\cite{RodriguezPuebla}.
This study is hence crucial for understanding the evolutionary path of SMGs and their link with local galaxies. 
Although a larger sample is needed to statistically estimate the properties of these sources and their contribution to the cosmic star formation history,
this galaxy suggests that star formation in dusty star-forming galaxies has been driven by similar physical processes during the last $\sim12.8$\ Gyr .

\begin{addendum}

 \item We thank Ian Smail for insightful comments that improved the quality of the paper. 
 JAZ acknowledges support from a mexican CONACyT studentship.
 RJI, LD and IO acknowledge support from ERC in the form of the Advanced Investigator Programme, 321302, COSMICISM.
 LD additionally acknowledges support from the ERC Consolidator Grant CosmicDust.
 HD acknowledges financial support from the Spanish Ministry of 
 Economy and Competitiveness (MINECO) under the 2014 Ram\'on y Cajal 
 program MINECO RYC-2014-15686.
 MJM acknowledges the support of the National Science Centre, Poland
through the POLONEZ grant 2015/19/P/ST9/04010 and  the European Union's Horizon 2020 research and innovation
programme under the Marie Sk{\l}odowska-Curie grant agreement No.
665778.
This work would not have been possible without the long-term
financial support from the Mexican CONACyT
during the construction and early operational phase of the
Large Millimeter Telescope {\it Alfonso Serrano}, as well as support
from the  US National Science Foundation via the University
Radio Observatory program, the Instituto Nacional de Astrof\'isica,
\'Optica y Electr\'onica (INAOE), and the University of Massachusetts
(UMass). The Submillimeter Array is a joint project between the Smithsonian 
Astrophysical Observatory and the Academia Sinica Institute of Astronomy and 
Astrophysics and is funded by the Smithsonian Institution and the Academia 
Sinica. ALMA is a partnership of ESO (representing its member states), NSF (USA) and NINS (Japan), 
together with NRC (Canada), MOST and ASIAA (Taiwan), and KASI (Republic of Korea), in 
cooperation with the Republic of Chile. The Joint ALMA Observatory is operated by ESO, AUI/NRAO and NAOJ.

\item[Author Contributions] JAZ led the scientific analysis and the writing of the paper, 
as well as the SMA follow-up proposal. RJI, EV, SE, AC, HD, JSD, LD, MJM, SS, IS, MWLS, and PW 
have contributed to the original {\it Herschel} proposals and source selection of the red sources, 
where this source was originally identify. AM, DHH, EV, IA, VAR, MC, DRG, ET, and OV  performed the selection of the 
sample for the LMT observations and lead the LMT proposals. MSY, GN, FPS, DS, GW, DSA, AV, and MZ carried out 
LMT data reduction and  interpretation. DW, MY, and AIGR assisted with the SMA observations and data reduction. 
JS, IO, HN have contributed to the data analysis and to fitting and modeling the results.
All the authors have discussed and contributed to this manuscript.

 \item[Correspondence] Correspondence and requests for materials
should be addressed to Jorge Zavala~(email: zavala@inaoep.mx).

\item[Competing Interests] The authors declare that they have no
competing financial interests.

\end{addendum}
\end{multicols}

\newpage
\begin{figure}
\begin{center}
\includegraphics[width=\textwidth]{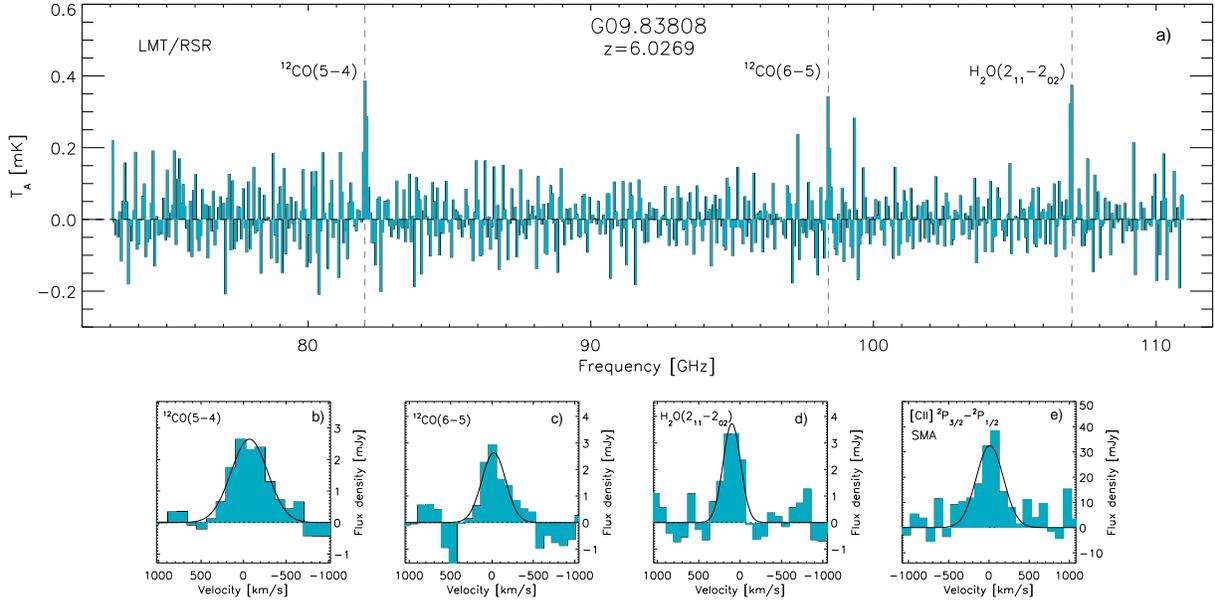}
\end{center}
\vspace{-0.3cm}
\caption{{\bf Identification of molecular emission  lines and redshift derivation. a),} Wide-band Redshift Search Receiver (RSR) 3\ mm spectrum of 
G09 83808 taken with the Large Millimeter Telescope (LMT). The transitions detected above $\rm S/N=5$ are marked with vertical dashed lines, 
and correspond to $^{12}$CO(5-4), $^{12}$CO(6-5), and H$_2$O($2_{11}-2_{02}$) at $z=6.0269\pm0.0006$. The spectrum has been rebinned into 2
pixels bins ($\sim200\rm km/s$) for better visualization. {\bf b), c), d),} LMT/RSR unbinned spectra at the position of the detected lines along with
the best-fitting Gaussian profiles. {\bf e),} SMA spectrum centered at the position of the detected line. The $x$-axes is in velocity offset with
respect to the derived redshift of $z=6.0269$. The derived properties of the lines are reported in Table 1.}
\end{figure}

\newpage
\begin{figure}
\begin{center}
\includegraphics[width=0.85\textwidth]{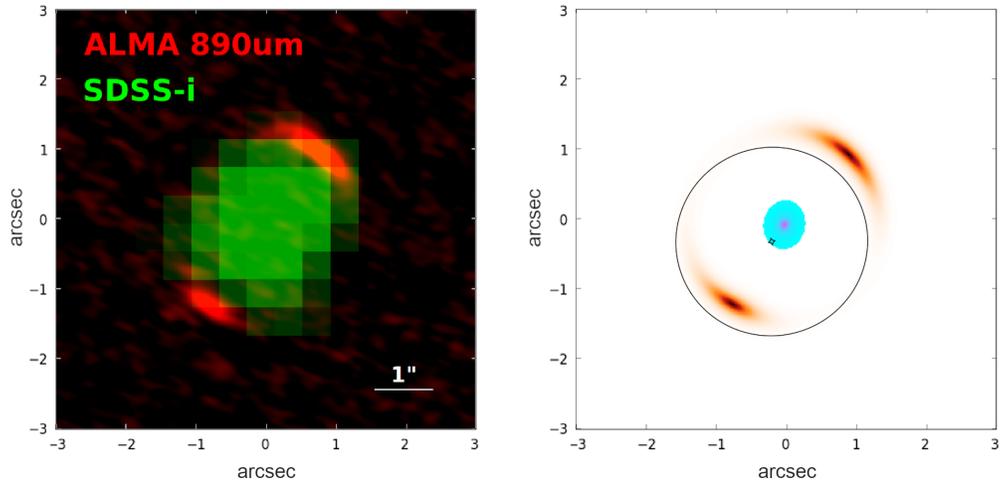}
\end{center}
\vspace{-0.3cm}
\caption{{\bf ALMA high-angular resolution continuum observations and lensing model.} {\it Left:} Color composite image of
G09 83808 centered at RA: $09^\text{h}00^\text{m}45.8$, Dec: $+00^\circ41'23''$. The green channel
represents the {\it i}-band data from SDSS and the red channel the ALMA\ 890\ $\mu\rm m$\ observations. An Einstein ring-like structure of radius  $\approx1.4$ arcsec
in the ALMA image is clearly seen around a foreground galaxy at $z=0.776$, which confirms that our high-redshift galaxy is strongly amplified. {\it Right:} 
Best-fit lensing model based on the visibilities of ALMA observations, from which we derived a gravitational amplification of $\mu=9.3\pm1.0$.}
\end{figure}

\newpage
\begin{figure}
\begin{center}
\includegraphics[width=0.7\textwidth]{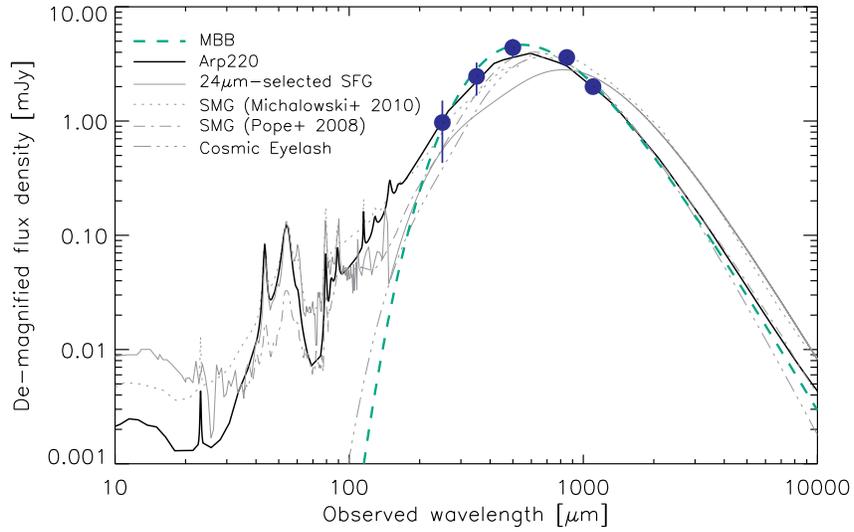}
\end{center}
\vspace{-0.3cm}
\caption{{\bf Photometry and spectral energy distribution (SED).} De-magnified (with $\mu=9.3\pm1.0$) flux densities at 250, 350, 500, 850 and 
1100\ $\mu$m from {\it Herschel/SPIRE}, JCMT/SCUBA-2, and LMT/AzTEC are indicated by the blue circles, with bars representing the photometric $1\sigma$ errors including calibration and lensing modeling uncertainties. These flux densities were fitted with different SED templates, including: Arp220, Cosmic Eyelash, two average SMG templates, an average 24\ $\mu$m-selected star-forming galaxy template, and a modified black body (MBB, see Methods section for details). We achieve the lowest $\chi^2$ with the Arp220 template, from which we derive an IR luminosity of $3.8\pm0.5\times10^{12}\rm\ L_\odot$ (corrected for magnification). From the best-fit modified black body distribution we derive a dust temperature of $49\pm3$\ K. As discussed in the Methods section, the CMB effects are not significant.}
\end{figure}

\newpage
\begin{figure}
\begin{center}
\includegraphics[width=0.5\textwidth]{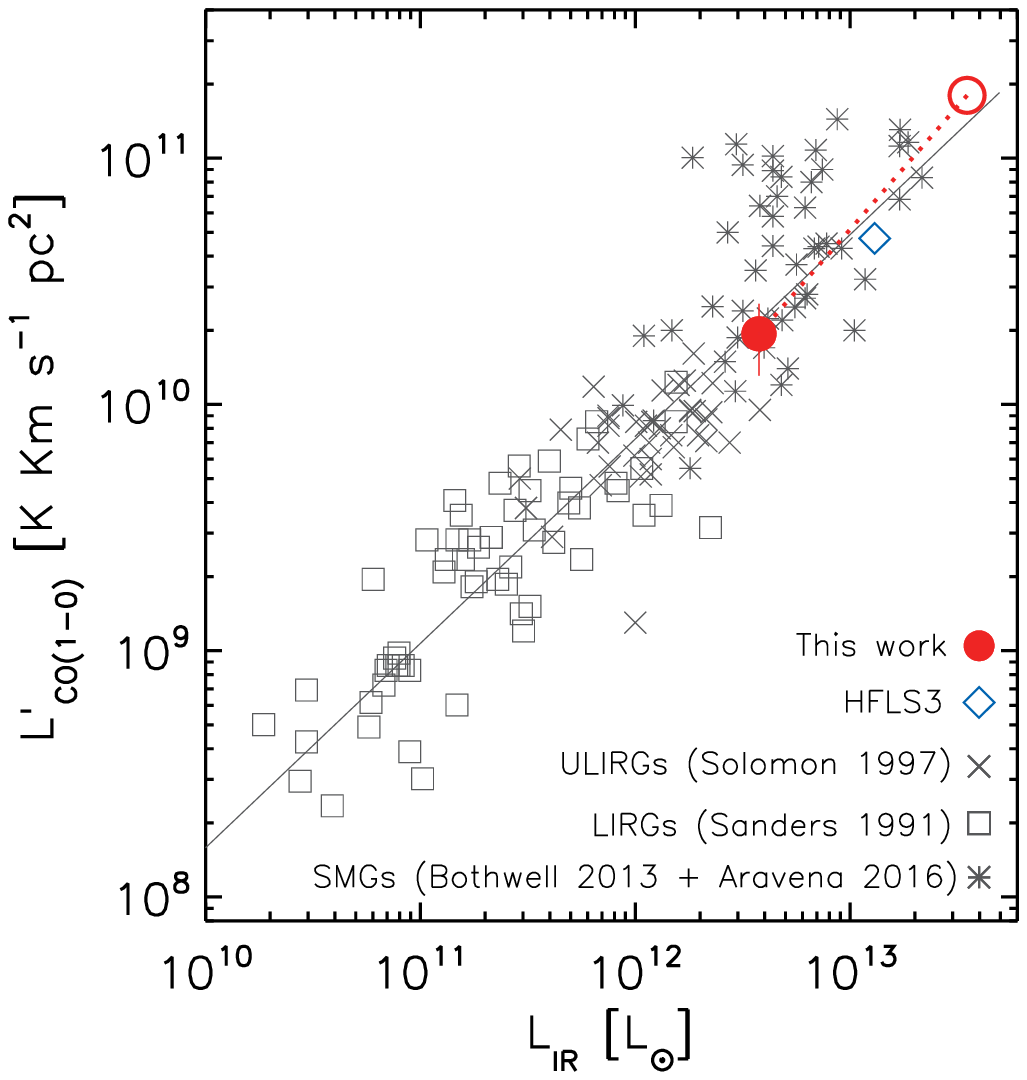}\includegraphics[width=0.5\textwidth]{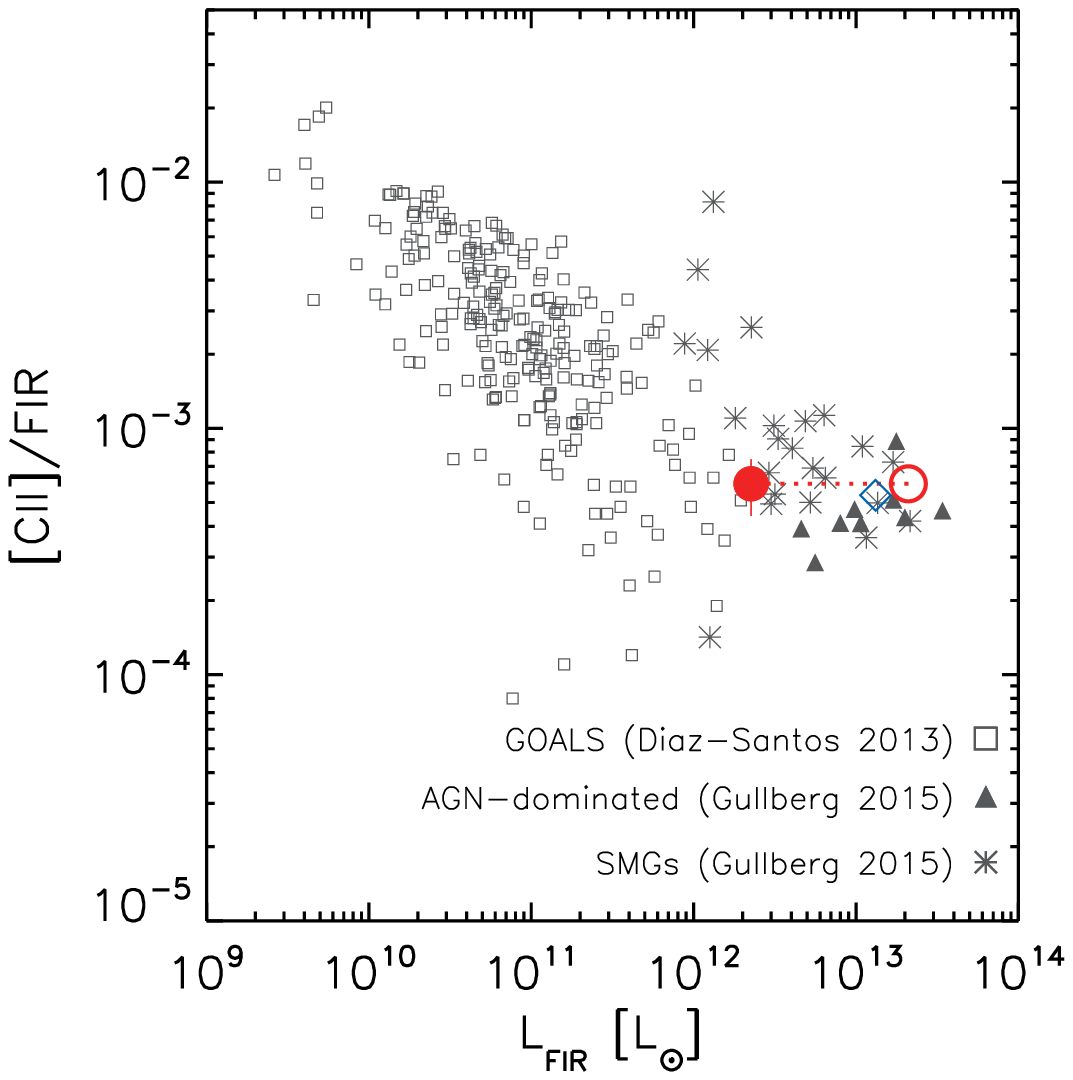}
\end{center}
\vspace{-0.3cm}
\caption{{\bf Star formation efficiency and [CII] deficiency.} {\it Left:} Lens-corrected CO(1-0) luminosity versus IR luminosity ($L'_{\rm CO(1-0)}-L_{\rm IR}$) 
as a proxy for the star-formation efficiency of G09 83808. For comparison, local LIRGS\cite{ref34}, ULIRGS\cite{ref27}, and lower-redshift SMGs\cite{ref28,aravena16} are 
plotted along with the best-fit relation to the three samples\cite{ref28}. As can be seen, G09 83808 falls on the same relation (as well as HFLS3\cite{ref11} after 
correcting for magnification\cite{cooray14}),
which suggests that the same star formation efficiency holds from $z\sim0$ to $z=6$ (i.e. during the last $\sim12.8$\ Gyr). 
The empty circle represents the position of our source if no lensing amplification correction is applied.
{\it Right:} [CII]/FIR versus de-magnified (filled circle) and amplified (empty circle) FIR luminosity for G09 83808. For comparison, we also plot a sample of
(U)LIRG galaxies from the Great Observatories All-sky Survey (GOALS\cite{ref30}), and a compilation of high-redshift sources\cite{ref35} that 
includes SMGs and AGN-dominated sources. As can be seen, our source follows the same trend found for local (U)LIRGs once corrected for magnification.}

\end{figure}

\begin{table*}[h]
\begin{center}
\small
\caption{{\bf Measured spectral line and continuum properties (not corrected for gravitational amplification).}  Photometric errors represent $1\sigma$ uncertainties in the flux density measurements including calibration errors. The $1\sigma$ uncertainties in the best-fitted Gaussian distribution parameters (central frequency, width, and integrated flux density) are also reported and propagated to estimate the error in the line luminosity.}
\begin{tabular}{lcccc|cc}
\hline
     &  \multicolumn{4}{c}{Transition}&\multicolumn{2}{c}{Photometry$^a$}\\
     & CO(5-4) & CO(6-5)  & H$_2$O($2_{11}-2_{02}$) &[CII] & [$\mu\rm m$]& [mJy]\\
\hline
$\nu_{\rm obs}$ [GHz] & $82.031\pm0.007$ & $98.41\pm0.01$ & $106.993\pm0.007$ & $270.35\pm0.03$ &250 &$9.7\pm5.4$\\
FWHM [$\rm km\ s^{-1}$]   & $490\pm60$         & $320\pm70$         & $240\pm40$         & $400\pm70$    &350 & $24.6\pm7.9$ \\
$S_{\rm int}$ [$\rm Jy\ km\ s^{-1}$]& $1.6\pm0.3$      & $0.9\pm0.3$      & $0.8\pm0.2$      & $13.8\pm3.0$  & 500& $44.0\pm8.2$ \\
$L'$ [$10^{10}\ \rm K\ km\ s^{-1}\ pc^{-2}$]& $7.6\pm1.2$ & $2.9\pm0.8$ & $2.3\pm0.5$  & $6.1\pm1.3$& 850&$36.0\pm3.1$\\
 & & & & & 1100&$20.0\pm1.0$ \\
\hline
\multicolumn{7}{l}{$^a$The flux densities at 250, 350, 500, and 850\ $\mu\rm m$ were taken from ref.\cite{ref12}}
\label{line_properties}
\end{tabular}
\end{center}
\end{table*}

\newpage

\begin{methods}

\begin{multicols}{2}
\section{Observations and data reduction}
\vspace{-0.4cm}
\subsection{LMT observations}

\vspace{-0.4cm}
\noindent Continuum and spectroscopic observations were obtained using the Large Millimeter Telescope (LMT\cite{ref13}, PI: D. Hughes), 
located on the summit of Volc\'an Sierra Negra 
({\it Tlilt\'epetl}), Mexico, at $\sim4600$ m.a.s.l. Observations were carried out during the Early Science Phase of the telescope using the 
1.1\ mm continuum camera, AzTEC\cite{ref54}, and the 3\ mm spectrograph, Redshift Search Receiver (RSR\cite{ref55}). During these observations only the inner
32-m diameter region of the telescope active surface was illuminated, which provided an effective beam size of $\approx8.5$\ arcsec
at 1.1\ mm and between $20-28$\ arcsec in the RSR 3\ mm window (75\ GHz - 110\ GHz). 

\vspace{-0.4cm}
\noindent AzTEC observations were performed on 2014 November 10 with an opacity of $\tau_{225}=0.07$ and total on-source integration time of 11\ min. Data reduction
were done following the AzTEC Standard Pipeline\cite{ref33}. G09 83808 was detected with a 
S/N\ $\approx20$ with a flux density of $S_{1.1\rm mm}=20.0\pm1.0$\ mJy. RSR observations were  subsequently taken at the AzTEC position  in 
two different periods: February 2016 and February 2017, along five different nights with an opacity range of $\tau_{225}=0.05-0.15$ and a total integration 
time of 8 hrs. Pointing observations on bright millimetre sources were done every hour. Data reduction was performed using the Data Reduction
and Analysis Methods in Python (DREAMPY). The final spectra were obtained by averaging all scans using $1/\sigma^2$ weights after flagging bad data. Finally,
to convert from antenna temperature units to flux, a factor of $7\rm\ Jy\ K^{-1}$ was used\cite{ref31}. 
The final spectrum shows three lines detected at 
S/N\ $\gtrsim5$ associated to CO(6-5), CO(5-4) and H$_2$O($2_{11}-2_{02}$) at $z=6.0269$. A cross-correlation template analysis\cite{ref31}
also identifies this redshift as the best solution with a S/N\ $=9.1$. 
Figure 1 shows the final spectrum after a Savitzky-Golay filter\cite{savitzky} has been applied for better visualization (the filter does not modified any of the properties 
of the detected lines).

\vspace{-0.4cm}
\noindent At the redshift of our source the [CII] 158\ $\mu$m line (see below) falls within the AzTEC band pass and then contributes to the total flux density
measured at 1.1\ mm. However, the contamination from the line is measured to be less than 2 per cent. Even if the [CII] line luminosity was as high as 1 per cent of 
the total IR luminosity, the contamination to the AzTEC measurement would be only $\sim6$ per cent, which is similar to the absolute flux calibration uncertainty. 
Therefore, and at least for this source, the contamination of the emission line to the  1.1\ mm continuum flux density is less important than anticipated\cite{ref32}.

\vspace{-0.2cm}
\subsection{SMA observations}

\vspace{-0.4cm}
\noindent G09 83808 was observed with the Submillimeter Array (SMA, PI: J. Zavala) on Mauna Kea,
Hawaii, on 2017 April 03. The weather conditions were good, with an average 
atmospheric opacity of $\tau_{225}=0.07$ and stable phase. The seven available
array antennas were in a compact configuration that provided baseline lengths
from 8 to 77 meters. The `345' receiver set was
tuned to provide spectral coverage $\pm(4-12)$~GHz from a LO frequency of
277.5~GHz, specifically to span a broad range around the estimated (redshifted)
[CII] line frequency of $\sim270.5$~GHz in the lower sideband. The SWARM
correlator provided uniform channel spacing of 140~kHz ($\sim$0.16~km~s$^{-1}$)
over the full bandwidth. The usable field of view is set by the FWHM primary
beam size of $\sim47$\ arcsec at this
frequency.

\vspace{-0.4cm}
\noindent The basic observing sequence consisted of a loop of 2 minutes each on the
gain calibrators J0909+013 (1.57 Jy) and J0831+044 (0.47 Jy) and
17.5 minutes on G09 83808. The track spanned an hour angle range of
$-0.8$ to $4.8$ for the target source. Passband calibration was obtained
with observations of the strong quasar 3C279. The absolute flux scale was
set using observations of Callisto, with an estimated accuracy of 20\%.
All of the basic calibration steps were performed using standard procedures
in the MIR software package. The calibrated visibilities were exported to
the {\sc MIRIAD} software package for imaging and deconvolution. Within {\sc MIRIAD},
the task {\tt uvaver} was used to combine the 4 correlator windows of the
lower sideband and to resample the visibilities to 50~km~s$^{-1}$ spectral
resolution. The task {\tt uvlin} was used for continuum subtraction, using
a linear fit to line-free channel ranges in the band. The task {\tt invert}
provided Fourier inversion for both continuum and spectral line imaging,
followed by {\tt clean} for deconvolution. The synthesized beam size obtained
with natural weighting was $2.5^{''}\times2.3^{''}$, p.a. $82^\circ$ for
the spectral line data cube, with rms noise 7.1~mJy per 50~km~s$^{-1}$ bin.
The final spectrum (see Fig.\ 1) was then extracted from a rectangular region that
comprise all the continuum emission. We measured the continuum flux density of the 
source to be $21.5\pm3$\ mJy, in very good agreement with the AzTEC photometry.

\vspace{-0.2cm}
\subsection{ALMA observations}

\vspace{-0.4cm}
\noindent The ALMA high-resolution $870 \, {\rm \mu m}$ observations used in this work were 
taken on 31 August 2015 (project 2013.1.00001.S, PI Rob Ivison; ref.\cite{oteo17} ), when the array was in a 
relatively extended configuration with baselines up to $1.6 \, {\rm km}$. The default
continuum spectral configuration was used, covering $[335.49 - 339.49]\, {\rm GHz}$ and 
$[347.49 - 351.49]\,{\rm GHz}$. The data were calibrated using the ALMA pipeline, with no 
further manual flagging required. The calibrated visibilities were imaged by using Briggs 
weighting with robust = 0.5, which is a good compromise between sensitivity and angular 
resolution. The beam size is then $\sim 0.12^{''}$ and the continuum sensitivity is 
${\rm r.m.s.} \sim 0.1 \, {\rm mJy \, beam^{-1}}$. The visibility weighting in ALMA data is
generally only correct in a relative sense, while our subsequent lens modeling procedure 
(see `Lensing Model' below) requires an absolute estimate of the noise in the data. The data weights
are determined by differencing successive visibilities on the same baseline, polarization,
and frequency baseband. The ALMA data also serendipitously cover the frequency of the 
redshifted $122 \, {\rm \mu m}$ [N{\scriptsize II}] line; this line is not detected at $>3\sigma$ significance.

\vspace{-0.2cm}
\section{Lensing model}

\vspace{-0.4cm}
\noindent The lens model was created using the publicly-available \texttt{visilens} code\cite{ref14}; details of the code 
are given in that work. Briefly, the lens mass profile is parameterized as a Singular Isothermal Ellipsoid, and the 
background source is modeled with a single elliptical S\'{e}rsic profile. The parameter space is explored using a 
Markov Chain Monte Carlo sampling method, generating a model lensed image at each proposed combination of lens and 
source parameters. The redshift of both sources is fixed at $z=0.776$ (based on X-Shooter/VLT observations\cite{fudamoto17}) and $z=6.027$, respectively. 
Because pixel values in interferometric images are correlated and subject to difficult-to-model
residual calibration errors, the proposed model image is inverted to the visibility domain and sampled at the $uv$ 
coordinates of the ALMA data. We also allow for 
residual antenna-based phase calibration errors in the model which 
could be due to, for example, uncompensated atmospheric delays. The phase shifts of all antennas are $<10$\,deg,
indicating that no significant antenna-based calibration problems remain. 

\vspace{-0.4cm}
\noindent The lensed emission is reasonably well-fit by a single background S\'{e}rsic component, leaving peak residuals of
$\sim4\sigma$ (the source is detected at peak significance $\sim20\sigma$). These residuals may indicate that either the lens,
source, or both are more complex than the simple parametric forms we have assumed. We have verified that an additional 
background source component is not statistically motivated. The best-fit magnification of the source is 
$\mu_\mathrm{890\mu m} = 9.3 \pm 1.0$, with an intrinsic flux density $S_\mathrm{890\ \mu\rm m} = 4.3 \pm 0.5$\ mJy and half-light
radius $0.10 \pm 0.01$'' ($=0.6\pm0.1$\ kpc). This compact morphology resembles the sizes found for local ULIRGs\cite{Lutz} ($\sim0.5$\ kpc), 
which are smaller than the typical values in SMGs ($\sim1.8$\ kpc, ref:\cite{hodge16}).

\vspace{-0.2cm}
\section{SED fitting and dust properties}

\vspace{-0.4cm}
\noindent We fit different galaxy SED templates to the photometry of G09 83808 through a $\chi^2$ minimization method.
We include the SED template of Arp220\cite{ref36}, Cosmic Eyelash\cite{ref37} (SMM J2135-0102), two average SMGs templates\cite{ref38,ref39},
and finally a composite SED of 24\ $\mu$m-selected star-forming galaxy\cite{ref40}. 
All the SED templates were fixed at $z=6.027$.
The Arp220 SED template gives us the best fit with $\chi_{\rm red}^2=0.7$. Using this template we derive an IR 
($8-1000\mu\rm m$) luminosity of $3.8\pm0.5\times10^{12}\rm\ L_\odot$ and a FIR ($42.5-122.5\mu\rm m$) luminosity of $2.3\pm0.3\times10^{12}\rm\ L_\odot$ 
(both corrected for gravitational amplification) . For comparison, if we adopt instead a SMGs average 
template ($\chi_{\rm red}^2=1.2$) we obtain $\rm L_{IR}=3.0\pm0.4\times10^{12}\rm\ L_\odot$, which is in good agreement with the value derived using 
the Arp220 template. Using Kennicutt standard relation\cite{ref41} for a Chabrier initial mass function (IMF)\cite{ref42}, this IR luminosity corresponds
to a star formation rate (SFR) of $380\pm50\rm\ M_\odot\ yr^{-1}$, or to $570\pm70\rm\ M_\odot\ yr^{-1}$ if the most recent relation\cite{Kennicutt2012} is used. 
If we adopt instead the Kennicutt calibration\cite{ref41} for a 
Salpeter IMF\cite{ref43}, the SFR increases to $640\pm90\rm\ M_\odot\ yr^{-1}$, still below the range probed by other SMGs at $z\gtrsim5$.

\vspace{-0.4cm}
\noindent We also use a modified blackbody function to fit our photometric measurements  described by
\begin{equation}
  S_\nu\propto \{1- \exp[-(\nu/\nu_0)^\beta]\}B(\nu,T_{\rm d}),
\end{equation}
where $S_\nu$ is the flux density at frequency $\nu$, $\nu_0$ is the rest-frame frequency at which the emission becomes optically thick,
$T_{\rm d}$ is the dust temperature,  $\beta$ is the emissivity index, and $B(\nu,T_{\rm d})$ is the Planck function at 
temperature $T_d$. To minimize the number of free parameters, the emissivity index is fixed (previous observational works 
suggest $\beta=1.5-2$; refs.\cite{ref44,ref45,ref46}), as well as $\nu_0=c/100\rm\ \mu m$
(refs.\cite{ref11,ref47}), where $c$ is the speed of light. From the best fit ($\chi^2\approx1.1$) we derive $T_d=49\pm3\rm\ K$ for 
$\beta=1.8$ and  $T_d=52\pm3\rm\ K$ for $\beta=1.5$. For these dust temperatures and at the redshift of our source the CMB effects\cite{ref48}
are not significant ($\Delta T\lesssim1\rm\ K$).

\vspace{-0.4cm}
\noindent Assuming the dust  is isothermal, the dust mass, $M_{\rm d}$, is estimated from 
\begin{equation}
M_{\rm d}=\frac{S_{\nu/(1+z)} D_L^2}{(1+z)\kappa_\nu B(\nu,T_{\rm d})},
\end{equation}
where $S_\nu$ is the  flux density at frequency $\nu$, $\kappa_\nu$ is the dust
mass absorption coefficient at $\nu$, $T_{\rm d}$ is the dust temperature,
and $B(\nu_,T_d)$ is the Planck function at temperature $T_{\rm d}$. The dust mass 
absorption follows the same power law as the optical depth, $\kappa\propto \nu^\beta$. 
Assuming normalization of $\kappa_{\rm d}(850\mu{\rm m})=0.07$ m$^2$ kg$^{-1}$ (ref.\cite{james2002}) and a dust temperature of $49\pm3\rm\ K$,
we estimate a dust mass of $\rm M_d=1.9\pm0.4\times10^8\ M_\odot$ after correcting for the CMB effects\cite{ref48} (although this
correction is less than 5 per cent). These calculations do not include the uncertainties of the dust mass absorption coefficient, 
which could be at least a factor of 3 (ref. \cite{ref49}). If we use instead a lower dust temperature of $35$\ K, the dust mass increases 
by a factor of $\sim2$.

\vspace{-0.4cm}
\noindent We also fit the observerd photometry with the {\sc MAGPHYS}\cite{cunha15} SED modelling code finding consistent results, within the error bars, 
with median values of SFR$=360^{+80}_{-70}\rm\ M_\odot\ yr^{-1}$, $\rm L_{IR}=4.5\pm0.7\times10^{12}\rm\ L_\odot$, $T_d=40^{+4}_{-2}\rm\ K$, and $\rm M_d=4.2\pm0.7\times10^8\ M_\odot$.

\vspace{-0.2cm}
\section{Spectral line properties}

\vspace{-0.4cm}
We calculate the line luminosity for each detected line following the standard relation\cite{ref50} described by:
\begin{equation}
L'_{\rm CO}=3.25\times 10^7 S_{CO}\Delta V\ \nu_{obs}^{-2}\ D^2_L\ (1+z)^{-3},
\end{equation}
where $L'_{\rm CO}$ is the line luminosity in K km s$^{-1}$ pc$^2$,  $S_{CO}\Delta V$ is the 
velocity-integrated line flux in Jy km s$^{-1}$, $\nu_{obs}$ is the observed central frequency 
of the line in GHz and $D_L$ is the luminosity distance in Mpc. The integrated flux, $S_{CO}\Delta V$,
is calculated as the integral of the best-fit Gaussian distribution, and its associated uncertainty 
through Monte Carlo simulations taking into account the errors in the Gaussian parameters (i.e. peak 
flux density and line width). To estimate the line luminosity in $\rm L_\odot$, we use 
$L=3\times10^{-11}\nu_r^3L'$, where $\nu_r$ is the rest frequency of the line\cite{ref50}. All properties are summarized in
Table 1.

\vspace{-0.2cm}
\section{CO(1-0) line luminosity and molecular gas mass}

\vspace{-0.4cm}
\noindent The molecular gas mass, $\rm M(H_2)$, can be derived using the CO luminosity to molecular gas mass conversion factor, $\alpha$, 
following the relation
\begin{equation}
\rm M(H_2)=\alpha\ L'_{CO(1-0)}.
\end{equation}

\vspace{-0.4cm}
\noindent For the  $\rm L'_{CO(1-0)}$ line luminosity we adopt the average value of $\rm L'_{CO(1-0)}=2.0\pm0.8\times10^{10}\ K\ km\ s^{-1}\ pc^{-2}$
extrapolated from our CO(6-5) and CO(5-4) transitions and correcting for gravitational amplification. The extrapolation was done 
using average brightness ratios found for lower-redshift SMGs\cite{ref28} ($\rm L'_{CO(5-4)}/L'_{CO(1-0)}=0.32\pm0.05$, $\rm L'_{CO(6-5)}/L'_{CO(1-0)}=0.21\pm0.04$),
this sample includes galaxies with similar luminosities to G09 83808 and are in agreement with those found for local ULIRGs\cite{greve14} (within the large scatter).
On the other hand, if we use the relationship between the Rayleigh-Jeans specific luminosity and CO(1-0) luminosity\cite{ref52}, 
$\rm L'_{CO(1-0)}\ [K\ km\ s^{-1}pc^2]=3.02\times10^{-21} L_{\nu}\ [\rm erg\ s^{-1}\ Hz^{-1}]$, we obtain a consistent line luminosity of $\rm 1\pm0.1\times10^{10}\ K\ km\ s^{-1}\ pc^{-2}$ (assuming 
a mass-weighted dust temperature of 35\ K, which is different from the luminosity-weighted dust temperature determined from the SED fitting\cite{ref52}).
Using the former value  and $\alpha=0.8\rm\ M_\odot\ (K\ km\ s^{-1}\ pc^{-2})^{-1}$,
which is appropriate for starburst galaxies\cite{ref51} (although some studies suggest larger values\cite{papadop12}),
we derive a molecular mass of $\rm M(H_2)=1.6\pm0.6\times10^{10}\rm\ M_\odot$. 

\vspace{-0.2cm}
\section{Dynamical mass and gas mass fraction}

\vspace{-0.4cm}
\noindent
Dynamical mass has been derived using the `isotropic virial estimator', which has been shown to be appropriate for lower-redshift SMGs\cite{engel10}:
\begin{equation}
M_{\rm dyn}[M_\odot]=2.8\times10^5\ \Delta\nu_{\rm FWHM}^2[{\rm km\ s^{-1}}]\ R_{1/2}[\rm kpc],
\end{equation}
where $\Delta\nu_{\rm FWHM}$ is the integrated line FWHM, which has been assumed to be 400\ km/s (as the average between the CO and [CII] lines), and $R_{1/2}$ 
is the half-light radius of $\sim0.6$\ kpc (derived from the lesing model of the continuum emission). This results in a dynamical mass of 
$M_{\rm dyn}=2.6\times10^{10}\rm\ M_\odot$. 
Using this estimation we calculate a gas mass fraction of $f_{\rm gas}=M_{\rm H_2}/M_{\rm dyn}\approx60$\%.
This constrain the CO luminosity to molecular gas mass conversion factor to $\alpha\lesssim1.4\rm\ M_\odot\ (K\ km\ s^{-1}\ pc^{-2})^{-1}$, otherwise the molecuar gas
mass would exceed the dynamical mass.

\newpage
\begin{addendum}
 \item[Data Availability] The datasets generated and analysed during this study are available from the corresponding 
 author on reasonable request. The ALMA observations presented here are part of the project 2013.1.00001.S  and the SMA
data correspond to the project 2016B-S078.
\end{addendum}

\noindent {\bf REFERENCES}\\

\end{multicols}

\end{methods}


\newpage

\noindent {\bf REFERENCES}

\begin{thebibliography}{1} 
{\small
\bibitem{ref53}Smail, I.  {\it et. al.} A Deep Sub-millimeter Survey of Lensing Clusters: A New Window on Galaxy Formation and Evolution.  {\it Astrophys. J. Lett.} {\bf 490} L5-L8 (1997). \vspace{-0.3cm}
\bibitem{ref2}Hughes, D. {\it et. al.} High-redshift star formation in the Hubble Deep Field revealed by a submillimetre-wavelength survey.  {\it Nature} {\bf 394}, 241-247 (1998). \vspace{-0.3cm}
\bibitem{ref3}Oliver, S. {\it et. al.} The Herschel Multi-tiered Extragalactic Survey: HerMES. {\it Mon. Not. R. Astron. Soc.} {\bf 424}, 1614-1635 (2012). \vspace{-0.3cm}
\bibitem{ref4}Valiante, E. {\it et. al.} The Herschel-ATLAS data release 1 - I. Maps, catalogues and number counts. {\it Mon. Not. R. Astron. Soc.} {\bf 462}, 3146-3179 (2016). \vspace{-0.3cm}
\bibitem{ref22}Micha{\l}owski, M. {\it et. al.} The SCUBA-2 Cosmology Legacy Survey: the nature of bright submm galaxies from 2 deg$^2$ of 850-$\mu$m imaging. {\it Mon. Not. R. Astron. Soc.} {\bf 469}, 492-515 (2017). \vspace{-0.3cm}
\bibitem{ref7}Capak, P. {\it et. al.} A massive protocluster of galaxies at a redshift of $z\sim5.3$. {\it Nature} {\bf 470} 233-235 (2011). \vspace{-0.3cm}
\bibitem{ref8}Combes, F. {\it et. al.} A bright z = 5.2 lensed submillimeter galaxy in the field of Abell 773. HLSJ091828.6+514223. {\it Astron. Astrophys. Lett.} {\bf 538}, L4 (2012). \vspace{-0.3cm}
\bibitem{ref9}Walter, F. {\it et. al.}  The intense starburst HDF850.1 in a galaxy overdensity at z=5.2 in the Hubble Deep Field. {\it Nature} {\bf 486}, 233-236 (2012). \vspace{-0.3cm}
\bibitem{ref10}Ma, J. {\it et. al.} Stellar Masses and Star Formation Rates of Lensed, Dusty, Star-forming Galaxies from the SPT Survey. {\it Astrophys. J.} {\bf812}, 88-104 (2015). \vspace{-0.3cm}
\bibitem{riechers17}Riechers, D. {\it et. al.} Rise of the Titans: A Dusty, Hyper-Luminous `870 micron riser' Galaxy at z $\sim$ 6.  {\it Astrophys. J.} arXiv preprint: 1705.09660 (2017). \vspace{-0.3cm}
\bibitem{ref11}Riechers, D. {\it et. al.} A Dust-Obscured Massive Maximum-Starburst Galaxy at a Redshift of 6.34. {\it Nature} {\bf 496}, 329-333 (2013). \vspace{-0.3cm}
\bibitem{ref57}Strandet, M. {\it et. al.} ISM properties of a Massive Dusty Star-Forming Galaxy discovered at z $\sim$ 7. {\it Astrophys. J. Lett.} {\bf 842} L15 (2017). \vspace{-0.3cm}
\bibitem{ref17}Sanders, D \& Mirabel, I. Luminous Infrared Galaxies. {\it Ann. Rev. Astron. Astrophys.} {\bf 34}, 749- (1996). \vspace{-0.3cm}
\bibitem{ref12}Ivison, R.  {\it et. al.} The Space Density of Luminous Dusty Star-forming Galaxies at $z> 4$: SCUBA-2 and LABOCA Imaging of Ultrared Galaxies from Herschel-ATLAS. {\it Astrophys. J.} {\bf 832}, 78- (2016). \vspace{-0.3cm}
\bibitem{cooray14}Cooray, A. {\it et. al.}  HerMES: The Rest-frame UV Emission and a Lensing Model for the z = 6.34 Luminous Dusty Starburst Galaxy HFLS3. {\it Astrophys. J.} {\bf 790}, 40- (2014). \vspace{-0.3cm}
\bibitem{capak2015}Capak, P. {\it et. al.} Galaxies at redshifts 5 to 6 with systematically low dust content and high [C II] emission {\it Nature} {\bf 522} 455-458 (2015). \vspace{-0.3cm}
\bibitem{ref18}Watson, D.  {\it et. al.} A dusty, normal galaxy in the epoch of reionization. {\it Nature} {\bf 519}, 327-330 (2015). \vspace{-0.3cm}
\bibitem{ref19}Willott, C., Carilli, C., Wagg, J. \& Wang, R. Star Formation and the interstellar medium in $z>6$  UV-luminous Lyman-break galaxies. {\it Astrophys. J.} {\bf 807}, 180- (2015). \vspace{-0.3cm}
\bibitem{ref58}Decarli, R. {\it et al.} Rapidly star-forming galaxies adjacent to quasars at redshifts exceeding 6. {\it Nature} {\bf 545}, 457-461 (2017). \vspace{-0.3cm}
\bibitem{ref56}Oteo, I. {\it et. al.} Witnessing the Birth of the Red Sequence: ALMA High-resolution Imaging of [C II] and Dust in Two Interacting Ultra-red Starbursts at z = 4.425. {\it Astrophys. J.} {\bf 827}, 34- (2016). \vspace{-0.3cm}
\bibitem{nayyeri}Nayyeri, H. {\it et. al.} A Study of Massive and Evolved Galaxies at High Redshift. {\it Astrophys. J.} {\bf 794}, 68- (2014). \vspace{-0.3cm}
\bibitem{ref24}Straatman, C. {\it et. al.} The Sizes of Massive Quiescent and Star-forming Galaxies at $z\sim4$ with ZFOURGE and CANDELS. {\it Astrophys. J. Lett.} {\bf 808}, L29- (2015). \vspace{-0.3cm}
\bibitem{ref28}Bothwell, M. {\it et. al.} A survey of molecular gas in luminous sub-millimetre galaxies. {\it  Mon. Not. R. Astron. Soc.} {\bf 429}, 3047-3067 (2013). \vspace{-0.3cm}
\bibitem{ref59}Tacconi, L. {\it et. al.} High molecular gas fractions in normal massive star forming galaxies in the young Universe. {\it Nature} {\bf 463}, 781-784 (2010).  \vspace{-0.3cm}
\bibitem{greve14}Greve, T. {\it et. al.} Star Formation Relations and CO Spectral Line Energy Distributions across the J-ladder and Redshift. {\it Astrophys. J.}  {\bf 794}, 142- (2014).\vspace{-0.3cm}
\bibitem{ref51}Carilli, C. \& Walter, F. Cool Gas in High-Redshift Galaxies. {\it Ann. Rev. Astron. Astrophys.}  {\bf 51}, 105-161 (2013). \vspace{-0.3cm}
\bibitem{yang2016}Yang, C.  {\it et. al.} Submillimeter H$_{2}$O and H$_{2}$O$^{+}$emission in lensed ultra- and hyper-luminous infrared galaxies at $z=2-4$.  {\it Astron. Astrophys.} {\bf 595}, 80- (2016).  \vspace{-0.3cm}
\bibitem{ref29}Wilson, C. {\it et. al.} Luminous Infrared Galaxies with the Submillimeter Array. I. Survey Overview and the Central Gas to Dust Ratio. {\it  Astrophys. J. Suppl. Ser.} {\bf 178}, 189-224 (2008). \vspace{-0.3cm}
\bibitem{ref30}D{\'{\i}}az-Santos, T. {\it et. al.} Explaining the [C II]157.7 {$\mu$}m Deficit in Luminous Infrared Galaxies - First Results from a Herschel/PACS Study of the GOALS Sample. {\it  Astrophys. J.} {\bf 774}, 68- (2013). \vspace{-0.3cm}
\bibitem{RodriguezPuebla}Rodriguez-Puebla, A., Primack, J., Avila-Reese, V. \& Faber, S. Constraining the galaxy-halo connection over the last 13.3 Gyr: star formation histories, galaxy mergers and structural properties. {\it Mon. Not. R. Astron. Soc.} {\bf 470}, 651-687 (2017). \vspace{-0.3cm}
}

\end{thebibliography}

\begin{thebibliography}{1} 
{\small
\makeatletter
\addtocounter{\@listctr}{30}
\makeatother
\bibitem{ref34}Sanders, D., Scoville, N. \& Soifer, B. Molecular gas in luminous infrared galaxies.  {\it Astrophys. J.} {\bf 370} 158-171 (1991). \vspace{-0.3cm}
\bibitem{ref27}Solomon, P. {\it et al.} The Molecular Interstellar Medium in Ultraluminous Infrared Galaxies. {\it Astrophys. J.} {\bf 478}, 144-161 (1997). \vspace{-0.3cm}
\bibitem{aravena16}Aravena, M. {\it et al.} A survey of the cold molecular gas in gravitationally lensed star-forming galaxies at z $>$ 2. {\it Mon. Not. R. Astron. Soc.} {\bf 457}, 4406-4420 (2016). \vspace{-0.3cm}
\bibitem{ref35}Gullberg, B.  {\it et al.} The nature of the [C II] emission in dusty star-forming galaxies from the SPT survey.  {\it Mon. Not. R. Astron. Soc.} {\bf 449} 2883-2900 (2015). \vspace{-0.3cm}
\bibitem{ref13}Hughes, D.  {\it et al.} The Large Millimeter Telescope. {\it SPIE} {\bf 7733}, 12- (2010). \vspace{-0.3cm}
\bibitem{ref54}Wilson, G. {\it et al.} The AzTEC mm-wavelength camera.  {\it Mon. Not. R. Astron. Soc.} {\bf 386} 807-818 (2008). \vspace{-0.3cm}
\bibitem{ref55}Erickson, N.  {\it et al.} An Ultra-Wideband Receiver and Spectrometer for 74--110 GHz  {\it ASPCS} {\bf 375} 71- (2007). \vspace{-0.3cm}
\bibitem{ref33}Scott, K. B. {\it et al.} AzTEC millimetre survey of the COSMOS field - I. Data reduction and source catalogue.  {\it Mon. Not. R. Astron. Soc. } {\bf 385} 12225-2238 (2008). \vspace{-0.3cm}
\bibitem{ref31}Yun, M. {\it et. al.} Early Science with the Large Millimeter Telescope: CO and [C II] Emission in the z = 4.3 AzTEC J095942.9+022938 (COSMOS AzTEC-1). {\it Mon. Not. R. Astron. Soc.} {\bf 454}, 3485-3499 (2015). \vspace{-0.3cm}
\bibitem{savitzky}Savitzky, A \& Golay, M. Smoothing and differentiation of data by simplified least squares procedures. {\it Analytical Chemistry}. {\bf 36} 1627-1639 (1964). \vspace{-0.3cm}
\bibitem{ref32}Smail, I. {\it et. al.} The potential influence of far-infrared emission lines on the selection of high-redshift galaxies. {\it Mon. Not. R. Astron. Soc. Lett.} {\bf 414}, L95-L99 (2011). \vspace{-0.3cm}
\bibitem{oteo17}Oteo, I. {\it et al.} Witnessing the birth of the red sequence: the physical scale and morphology of dust emission in hyper-luminous starbursts in the early Universe. ArXiv preprint: 1709.04191 (2017)
\bibitem{ref14}Spilker, J.  {\it et. al.} ALMA Imaging and Gravitational Lens Models of South Pole Telescope-Selected Dusty, Star-Forming Galaxies at High Redshifts. {\it Astrophys. J.} {\bf 826}, 112- (2016). \vspace{-0.3cm}
\bibitem{fudamoto17}Fudamoto, Y. {\it et al.} The most distant, luminous, dusty star-forming galaxies: redshifts from NOEMA and ALMA spectral scans. {\it Mon. Not. R. Astron. Soc.} accepted (2017). \vspace{-0.3cm}
\bibitem{Lutz}Lutz, D. {\it et al.} The far-infrared emitting region in local galaxies and QSOs: Size and scaling relations. {\it  Astron. Astrophys.} {\bf 591} 136- (2016). \vspace{-0.3cm}
\bibitem{hodge16}Hodge, J. {\it et al.} Kiloparsec-scale Dust Disks in High-redshift Luminous Submillimeter Galaxies. {\it Astrophys. J. } {\bf 833} 103- (2016). \vspace{-0.3cm}
\bibitem{ref36}Silva, L.  {\it et al.} Modeling the Effects of Dust on Galactic Spectral Energy Distributions from the Ultraviolet to the Millimeter Band.  {\it Astrophys. J. } {\bf 509} 103-117 (1998). \vspace{-0.3cm}
\bibitem{ref37}Ivison, R.  {\it et al.} {\it Herschel} and SCUBA-2 imaging and spectroscopy of a bright, lensed submillimetre galaxy at z = 2.3.  {\it Astron. Astrophys. Lett.} {\bf 518} L35- (2010). \vspace{-0.3cm}
\bibitem{ref38}{Micha{\l}owski}, M., Hjorth, J. \& Watson, D. Cosmic evolution of submillimeter galaxies and their contribution to stellar mass assembly. {\it Astron. Astrophys.} {\bf 514}, A67- (2010).  \vspace{-0.3cm}
\bibitem{ref39}Pope, A. {\it et al.} Mid-Infrared Spectral Diagnosis of Submillimeter Galaxies. {\it Astrophys. J. } {\bf 675}, 1171-1193 (2008).  \vspace{-0.3cm}
\bibitem{ref40}Kirkpatrick, A. {\it et al.} GOODS-{\it Herschel}: Impact of Active Galactic Nuclei and Star Formation Activity on Infrared Spectral Energy Distributions at High Redshift. {\it Astrophys. J. } {\bf 759}, 139- (2012).  \vspace{-0.3cm}
\bibitem{ref41}Kennicutt, R. The Global Schmidt Law in Star-forming Galaxies. {\it Astrophys. J. }  {\bf  498}, 541-552 (1998).\vspace{-0.3cm}
\bibitem{ref42}Chabrier, G. The Galactic Disk Mass Function:  Reconciliation of the Hubble Space Telescope and  Nearby Determinations.  {\it Astrophys. J. } {\bf 586}, L133-L136 (2003).\vspace{-0.3cm}
\bibitem{Kennicutt2012} Kennicutt, R. \&  Evans, N. Star Formation in the Milky Way and Nearby Galaxies. {\it Ann. Rev. Astron. Astrophys.} {\bf 50}, 531-608 (2012).\vspace{-0.3cm}
\bibitem{ref43}Salpeter, E.  The luminosity function and stellar evolution. {\it Astrophys. J. } {\bf 121}, 161- (1955).\vspace{-0.3cm}
\bibitem{ref44}Dunne, L. \& Eales, S. The SCUBA Local Universe Galaxy Survey - II. 450-{$\mu$}m data: evidence for cold dust in bright IRAS galaxies. {\it   Mon. Not. R. Astron. Soc. Lett.} {\bf 327}, 697-714 (2001).\vspace{-0.3cm}
\bibitem{ref45}Chapin, E. {\it et al.} An AzTEC 1.1mm survey of the GOODS-N field - II. Multiwavelength identifications and redshift distribution. {\it   Mon. Not. R. Astron. Soc. Lett.}  {\bf 398}, 1793-1808 (2009).\vspace{-0.3cm}
\bibitem{ref46}Magnelli, B. {\it et al.} A {\it Herschel} view of the far-infrared properties of submillimetre galaxies. {\it  Astron. Astrophys. }  {\bf 539}, 155- (2012).\vspace{-0.3cm}
\bibitem{ref47}Simpson, J. {\it et al.} The SCUBA-2 Cosmology Legacy Survey: Multi-wavelength Properties of ALMA-identified Submillimeter Galaxies in UKIDSS UDS. {\it Astrophys. J.}  {\bf 839}, 58- (2017).\vspace{-0.3cm}
\bibitem{ref48}da Cunha, E. {\it et al.} On the Effect of the Cosmic Microwave Background in High-redshift (Sub-)millimeter Observations. {\it Astrophys. J.}  {\bf 766}, 13- (2013).\vspace{-0.3cm}
\bibitem{james2002}James, A., Dunne, L., Eales, S. \& Edmunds, M. SCUBA observations of galaxies with metallicity measurements: a new method for determining the relation between submillimetre luminosity and dust mass. {\it   Mon. Not. R. Astron. Soc. Lett.} {\bf 335}, 753-761 (2002).\vspace{-0.3cm}
\bibitem{ref49}Dunne, L. {\it et al.} Type II supernovae as a significant source of interstellar dust. {\it Nature}  {\bf 424}, 285-287 (2003).\vspace{-0.3cm}
\bibitem{cunha15}da Cunha, E.  {\it et al.} An ALMA Survey of Sub-millimeter Galaxies in the Extended Chandra Deep Field South: Physical Properties Derived from Ultraviolet-to-radio Modeling. {\it Astrophys. J.}  {\bf 806}, 110- (2015).\vspace{-0.3cm}
\bibitem{ref50}Solomon, P. \& Vanden Bout, P. Molecular Gas at High Redshift. {\it Ann. Rev. Astron. Astrophys.}  {\bf 43}, 677-725 (2005).\vspace{-0.3cm}
\bibitem{ref52}Scoville, N. {\it et al.} ISM Masses and the Star formation Law at $z = 1$ to 6: ALMA Observations of Dust Continuum in 145 Galaxies in the COSMOS Survey Field. {\it Astrophys. J.}  {\bf 820}, 83- (2016).\vspace{-0.3cm}
\bibitem{papadop12}Papadopoulos, P. {\it et al.} The Molecular Gas in Luminous Infrared Galaxies. II. Extreme Physical Conditions and Their Effects on the X $_{co}$ Factor.  {\it Astrophys. J.} {\bf 751}, 10- (2012).\vspace{-0.3cm}
\bibitem{engel10}Engel, H. {\it et al.} Most Submillimeter Galaxies are Major Mergers.  {\it Astrophys. J.} {\bf 724}, 233-243 (2010).\vspace{-0.3cm}
}

\end{thebibliography}
\end{document}